\documentclass[]{raa}            
\usepackage{graphicx,times}
\usepackage{natbib}
\usepackage{txfonts}
\usepackage{dcolumn}
\input{psfig.sty}
\providecommand{\bm}[1]{\mbox{\boldmath$#1$\unboldmath}}

\makeatletter

\newcommand{\Rmnum}[1]{\expandafter\@slowromancap\romannumeral #1@}

\begin{document}

\title{Constraints on smoothness parameter and dark energy using observational $H(z)$ data}


   \author{Hao-Ran Yu
      \inst{1}
   \and Tian Lan
      \inst{1}
   \and Hao-Yi Wan
      \inst{3}
   \and Tong-Jie Zhang
      \inst{1,5}
   \and Bao-Quan Wang
      \inst{2}
   }

   \institute{Department of Astronomy, Beijing Normal University, 100875, Beijing, P. R. China;
        \and
            Department of Physics, Dezhou University, Dezhou 253023,
            P. R. China;
        \and
             Business Office, Beijing Planetarium, No. 138 Xizhimenwai Street, Beijing 100044,
             P. R. China;
        \and
             Center for High Energy Physics, Peking University, Beijing 100871, P. R. China \\{\it tjzhang@bnu.edu.cn}
}

\abstract{ The universe, with large-scale homogeneity, is locally inhomogeneous, clustering into stars, galaxies and larger structures. Such
property is described by the smoothness parameter $\alpha$ which is defined as the proportion of matter in the form of intergalactic medium. If
we take consideration of the inhomogeneities in small scale, there should be modifications of the cosmological distances compared to a
homogenous model. Dyer and Roeder developed a second-order ordinary differential equation (D-R equation) that describes the angular diameter
distance-redshift relation for inhomogeneous cosmological models. Furthermore, we may obtain the D-R equation for observational $H(z)$ data
(OHD). The density-parameter $\Omega_{\rm M}$, the state of dark energy $\omega$, and the smoothness-parameter $\alpha$ are constrained by a set
of OHD in a spatially flat $\Lambda$CDM universe as well as a spatially flat XCDM universe. By using of $\chi^2$ minimization method we get
$\alpha=0.81^{+0.19}_{-0.20}$ and $\Omega_{\rm M}=0.32^{+0.12}_{-0.06}$ at $1\sigma$ confidence level. If we assume a Gaussian prior of
$\Omega_{\rm M}=0.26\pm0.1$, we get $\alpha=0.93^{+0.07}_{-0.19}$ and $\Omega_{\rm M}=0.31^{+0.06}_{-0.05}$. For XCDM model, $\alpha$ is
constrained to $\alpha\geq0.80$ but $\omega$ is weakly constrained around -1, where $\omega$ describes the equation of the state of the dark
energy ($p_{\rm X}=\omega\rho_{\rm X}$). We conclude that OHD constrains the smoothness parameter more effectively than the data of SNe Ia and
compact radio sources.  }

  \authorrunning{Hao-Ran Yu, Tian Lan, Shi-Yu Li, Hao-Yi Wan, Tong-Jie Zhang \and Bao-Quan Wang}   
  \titlerunning{Constraints on smoothness parameter and dark energy using observational $H(z)$ data}   
  \maketitle

%
%
\section{Introduction}           
\label{sect:intro} Through recent several decades, we have entered into an accurate cosmological period and understood our universe more deeply.
According to the cosmological principle, our universe is homogeneous and isotropic on large scale, but the deviations from the homogeneities are
also observed. The corresponding researches are still exciting.

Recently, there is mounting data from type Ia supernovae, cosmic microwave background (CMB) and the large scale structure suggest that the
present universe is spatially flat and accelerated expanding. Combined analysis of the above cosmological observations support that an
approximately 26\% of cold dark matter (CDM) and the other part 74\% dominated by an unknown exotic component with negative pressure---the
so-called dark energy---driving the current acceleration
(\cite{perm1998};\cite{perm1999};\cite{Riess};\cite{Riess07};\cite{Ef02};\cite{Allen04};\cite{Astier06};\cite{Spergel02}). The most likely
candidate of this component is the cosmological constant (\cite{Carrllo}). In addition, dynamical models like quintessence (\cite{Caldwell98}),
 Chaplygin Gas (\cite{Kam00}),  "X-matter" model (\cite{turner97};\cite{Chiba97};\cite{Alcaniz99};\cite{Alcaniz01};\cite{Lima00};\cite{Lima03};\cite{browski07}),
Braneworld models (\cite{Csaki}) and the Cardassian models (\cite{Freese}) and soon proposed to explain the accelerating expansion of the
universe. In the case of X-matter, the dark energy has the following property with an equation of state:
\begin{equation}
    p_{\rm X}=\omega\rho_{\rm X},
    \label{eq:state}
\end{equation}
where $\omega$ is a constant independent of time or redshift. If $\omega=-1$, it is reduced to the case of the cosmological constant (the
$\Lambda$CDM model).

However, these models could not explain the observations of our universe perfectly. Except for the cosmological constant problem
(\cite{Weinberg89}), the deviation from the cosmological principle---the universe is homogenous and isotropic in large scale---needed to be
considered. It is obvious that the matter in the universe is clustered into stars, galaxies and clusters of galaxies, rather than absolutely
distributes uniformly and disperses everywhere in the space. It is also well known that the universe are grouped into superclusters, or perhaps
filaments, great walls and voids in larger scale. Only on the scale larger than 1 Gpc does the universe appear smooth. Such problems may have
effects on the distance redshift relation. Therefore a smoothness-parameter $\alpha$ was introduced to describe the proportion of the mean
density $\rho$ in the form of intergalactic matter (\cite{DR73}):
\begin{equation}
    \alpha\equiv\frac{\rho_{\rm int}}{\rho},
    \label{eq:defalpha}
\end{equation}
where $\rho_{\rm int}$ is the mean density in the universe in the form of intergalactic matter, while $\rho$ denotes the mean density of the
whole universe, so $\alpha\in[0,1]$. In the case of $\alpha=0$, it describes a universe with all the matter clustered into stars, galaxies and
so on. while $\alpha=1$, it is a normal homogeneous universe. Generally, $0<\alpha<1$ describes the universe partially in the form of clustered
matter and the other in the form of intergalactic matter.

The properties of angular diameter distance in a locally inhomogeneous universe has been discussed in Ref.
(\cite{Weinberg89};\cite{Zel67};\cite{Dashv66};\cite{Kayser97}). Later Dyer {\it et al.} established the Dyer-Roeder (D-R) equation to explain
the distance-redshift relation in a universe with fractional intergalactic medium (\cite{DR73}), as well as without intergalactic medium
(\cite{DR72}). In the literature (\cite{SantosSN}), by use of two different samples of SNe type Ia data, the $\Omega_{\rm M}$ and $\alpha$
parameters are constrained by $\chi^{2}$ minimization fitting, which applies the Zeldovich-Kantowski-Dyer-Roeder (ZKDR) luminosity distance
redshift relation for a flat $\Lambda$CDM model. A $\chi^{2}$-analysis, by using the 115 SNe Ia data of Astier {\it et al.} sample
(\cite{Astier}), constrains the density-parameter to be $\Omega_{\rm M}=0.26_{-0.07}^{+0.17}$ ($2\sigma$) while  the $\alpha$ parameter is
unlimited (all the values $\alpha\in  [0,1]$ are allowed even at 1$\sigma$). And, the analysis based on the 182 SNe Ia data of Riess {\it et
al.} (\cite{Riess2}) constrains the pair of parameters to be $\Omega_{\rm M}= 0.33^{+0.09}_{-0.07}$ and $\alpha\geq 0.42$ ($2\sigma$), which
provides a more stringent constraint because the sample extends to higher redshifts.

In Ref.(\cite{SantosR}), Santos {\it et al.} has proposed to constrain $\alpha$, $\Omega_{\rm M}$ and $\omega$ by the angular diameter distances
of compact radio sources with XCDM model. Howerver, only are the $\omega$ and $\Omega_{\rm M}$ parameters well constrained, but the $\alpha$
parameter is totally unconstrained in $1\sigma$ level.

As can be seen, neither SNe Ia nor compact radio sources data are capable of constraining the smoothness parameter. We could make use of other
astronomical data to constrain the smoothness parameter. It is also feasible to constrain inhomogeneous model by making use of the observational
$H(z)$ data (OHD), which can be obtained by the method to estimate the differential ages of the oldest galaxies. In Ref.(\cite{Yi}), Ze-Long Yi
and Tong-Jie Zhang present a constraint on a flat FRW universe with a matter component and a holographic dark energy component use OHD. In
Ref.(\cite{Wan}), Hao-Yi Wan {\it et al.} use OHD to constrain the Dvali Gabadadze Porrati (DGP) Universe. In Ref.(\cite{Lin}), Hui Lin {\it et
al.} successfully use OHD together with other observational data to constrain the $\Lambda$CDM cosmology. The wiggling Hubble parameter $H(z)$
are also studied (\cite{Zhanghongsheng}).

It can be conclude that OHD is a complementarity to other cosmological probes and may also present better constraint on the smoothness
parameter. In this article, the parameters $\Omega_{\rm M}$, $\alpha$ and $\omega$ are constrained by totally 12 bins of OHD from Simon {\it et
al.} (2005) and Ruth {\it et al.} (2008) in spatially flat $\Lambda$CDM universes as well as in the XCDM model. This paper is organized as
follows: In Sec.\ref{sec:theory}, we review the basic origin of Dyer-Roeder Equation and the relationship between Hubble parameter and different
cosmological models, which is performed in a inhomogeneous universe. In Sec.\ref{sec:results}, we constrain the parameters $\Omega_{\rm M}$,
$\omega$ and $\alpha$ from OHD. Discussions and prospects are presented in the third section.

\section{Dyer-Roeder Equation and the relationship
between ZKDR distance and Hubble parameter}\label{sec:theory}

We consider a stellar object which emits a beam of light propagating throughout a space-time described by the metric tensor $g_{\mu \nu}$. We
can identify a null surface $\Sigma$ determined by the eikonal equation $g^{\mu\nu}\Sigma_{,\mu}\Sigma_{,\nu}=0$ along which the beam of light
propagates. The direction of this light is the tangent vector of null surface i.e. the null geodesic $k_{\mu}=-\Sigma_{,\mu}$. The beam of light
rays can be described by $x^{\mu}=(v,y^i)$, where $x^0=\nu$ is the affine parameter and $y^i$ ($i=1,2,3$) indicates the three different
directions of the propagation of the light. The vector field is tangent to the light ray congruence, $k^{\mu}=\displaystyle{\frac{{\rm
d}x^{\mu}}{{\rm d}v}=-\Sigma,_{\mu}}$, determines two optical scalars---$\theta$ describing the convergence of the light and the shear parameter
$\sigma$,
\begin{equation}
\theta\equiv\frac{1}{2}{k^{\mu}}_{;\mu},\; \ \ \ \ \sigma\equiv k_{\mu;\nu}\tilde{m}^{\mu}\tilde{m}^{\nu},
\end{equation}
where $\tilde{m}^{\mu}=\displaystyle{\frac{1}{\sqrt{2}}}(\xi^{\mu}-i\eta)$ is a complex vector that is orthogonal to $k^{\mu}$
($k^{\mu}\tilde{m}_{\mu}=0$). Since $k_{\mu}=-\Sigma,_{\mu}$, the vorticity which is connected with the light beam is zero, therefore the
congruence of light is characterized by these two optical scalars, $\sigma$ and $\theta$. And these two optical scalars satisfy the Sachs
propagation equations (\cite{Sachs66}):
\begin{equation}
    \dot{\theta} +\theta^2 + |\sigma |^2 = - \frac{1}{2} {\mathcal R}_{\mu\nu}
    k^{\mu}k^{\nu},
    \label{eq:propagation1}
\end{equation}
\begin{equation}
    \dot{\sigma} + 2\theta \sigma = - \frac{1}{2} {\mathcal C}_{\mu\nu\tau\lambda}
    {\tilde{m}}^{\mu} k^{\nu}{\tilde{m}}^{\tau} k^{\lambda},
    \label{eq:propagation2}
\end{equation}
where a dot denotes the derivative with respect to $v$, ${\mathcal R}_{\mu\nu}$ and ${\mathcal R}$ are the Ricci tensor and Ricci scalar
respectively, and ${\mathcal C}_{\mu\nu\tau\lambda}$ is the Weyl tensor which is zero in a conformally flat FRW space-time. One can see that if
the shear $\sigma$ is initially zero, the Weyl tensor could always be zero, which automatically satisfies the condition in FRW space-time
(\cite{Demianski}). Therefore, assuming that the light beam has no shear --- $\sigma=0$, we may describe the convergence and divergence of this
beam of light by the parameter $\theta$(empty beam approximation).

\begin{figure}
\centerline{\psfig{figure=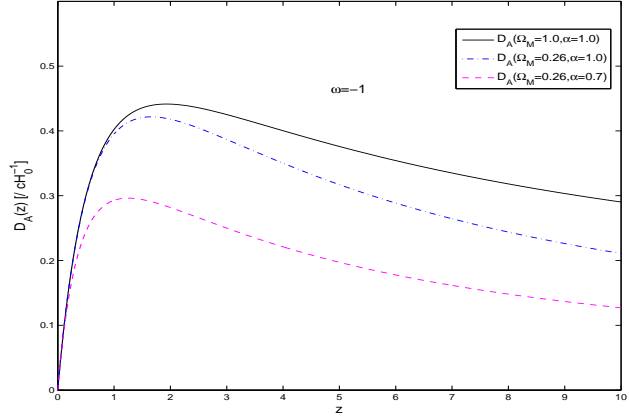,width=3.8truein,height=2.5truein} \hskip 0.0in} \caption{(color online). Angular diameter distance $D_A(z)$ as
a function of redshift $z$ for a flat $\Lambda$CDM model. Several selected values of $\Omega_{\rm M}$ and $\alpha$ are shown. $D_A(z)$ is in
units of $c/H_0$.} \label{fig:ang}
\end{figure}
The relative rate of the change of an infinitesimal area $A$ on the cross section of the beam can be described by the optical scalar $\theta$
(and the distortion by $\sigma$), the only parameter that characterizes the congruence of light, which relates to $A$ by
\begin{equation}
\theta=\frac{1}{2}\frac{\dot{A}}{A}.
\end{equation}
Substituting Eq. (\ref{eq:propagation1}) into the above expression, one can reduce the optical scalar equation to (\cite{Sachs61})
\begin{equation}
    \ddot{\sqrt{A}}+\frac{1}{2}{{\mathcal R}_{\mu\nu}k^{\mu}k^{\nu}\sqrt{A}}=0.
    \label{eq:Sachs}
\end{equation}
The Einstein field equation is
\begin{equation}
{\mathcal R}_{\mu\nu}-\frac{1}{2}g_{\mu\nu}{\mathcal R}-\lambda g_{\mu\nu}=-\frac{8\pi G}{c^2}T_{\mu\nu},
\end{equation}
multiply each side by $k^{\mu}k^{\nu}$, the two $g_{\mu\nu}$ terms vanish while leaving the following form,
\begin{equation}
    {\mathcal R}_{\mu\nu}k^{\mu}k^{\nu}=-\frac{8\pi
    G}{c^2}T_{\mu\nu}k^{\mu}k^{\nu}.
    \label{eq:field1}
\end{equation}
The universe, though locally inhomogeously distributed, is homogeneous and isotropic on the largest scale in the mean, so we choose
Robertson-Walker metric
\begin{equation}
{\rm d}s^2=c^2{\rm d}t^2-a^2(t){\rm d}\sigma^2,
\end{equation}
where ${\rm d}\sigma^2$ describes the spacial part of the metric and $a(t)$ is the scale factor of the universe. We set $a(t)=1$ at the present
time and choose proper affine parameter $v$ so that, (\cite{Sch56})
\begin{equation}
\frac{{\rm d}t}{{\rm d}\tau}=\frac{a_0}{H_0a},
\end{equation}
where $a_0$ and $H_0$ are the present values of $a$ and the Hubble constant, respectively. Then we have
\begin{equation}
k^0=\frac{{\rm d}x^0}{{\rm d}\tau}=\frac{{\rm d}(ct)}{{\rm d}\tau}=\frac{ca_0}{H_0a}.
\end{equation}
We consider a pressureless matter dominant universe in which the energy-momentum tensor has only a nonzero 0-0 part (with comoving coordinates),
i.e. $T_{00}=\rho$ and $T_{ik}=0$. In addition, if $a/a_0=(1+z)^{-1}$ and $\rho/\rho_0=(1+z)^3$, we get
\begin{equation}
T_{\mu\nu}k^{\mu}k^{\nu}=\left(\frac{ca_0}{H_0a}\right)^2\alpha\rho =\frac{c^2}{H_0^2}(1+z)^5\alpha\rho_0.
\end{equation}
Substituting it into Eq. (\ref{eq:field1}) and then substituting the resulting equation into Eq. (\ref{eq:Sachs}), we can finally obtain
\begin{equation}
\ddot{\sqrt{A}}+\frac{2}{3}\alpha\Omega_{\rm M}(1+z)^5\sqrt{A}=0. \label{eq:pre-DR}
\end{equation}
Here, we use the density-parameter $\Omega_{\rm M}$ instead of $\rho_0$. Due to a relationship between the angular diameter distance $D_{\rm A}$
and $A$, $D_{\rm A}=\sqrt{A}$ (\cite{Schneider};\cite{Schneider88};\cite{Bartelmann91};\cite{Watanabe92}), the Eq. (\ref{eq:pre-DR}) becomes the
Dyer-Roeder Equation (\cite{DR73}):
\begin{equation}
\ddot{D_{\rm A}}+\frac{2}{3}\alpha\Omega_{\rm M}(1+z)^5D_{\rm A}=0. \label{eq:DR}
\end{equation}
where the dots denote the derivatives with respect to affine parameter $v$.

It is necessary to mention that the smoothness-parameter must be different at various epoch of the universe due to the theory of formation of
the large scale structure (\cite{SantosR};\cite{Ef02}). For very high redshift, the matter in the universe must be more smoothly distributed
compared to that of present. In this point of view, we would have to identify the smoothness-parameter $\alpha$ as a function of $z$,
$\alpha(z)$ in Eq. (\ref{eq:DR}), especially when discussing the properties of the angular diameter distance at high redshift. However, because
the samples of compact radio sources, SNe Ia and OHD are mostly located at low redshifts ($z<2$ for Hubble parameter), we set $\alpha$ as a
constant in the following discussion. We will not consider the variations of $\alpha$ with respect to $z$ also because the data are neither
adequate nor precise enough . The redshift dependence of $\alpha$ was discussed by Santos {\it et al.} in Ref. (\cite{SantosR}).

\begin{table}
\begin{center}
\begin{tabular}{cccc}
\hline\hline \
\ \ Redshift $z$ & \ \ \ $H(z)$ \ \ \ & $1\sigma$ interval \ \ \ \ & \ \ \ Data \ \ \\
\hline
0.05 & 75.4 & $\pm 2.3$ & $\vardiamondsuit$  \\
0.09 & 69 & $\pm 12$ & $\star$ \\
0.17 & 83 & $\pm 8.3$ & $\star$ \\
0.27 & 70 & $\pm 14$ & $\star$ \\
0.40 & 87 & $\pm 17.4$ & $\star$ \\
0.505 & 96.9 & $\pm 6.9$ & $\vardiamondsuit$ \\
0.88 & 117 & $\pm 23.4$ & $\star$ \\
0.905 & 116.9 & $\pm 11.5$ & $\vardiamondsuit$ \\
1.30 & 168 & $\pm 13.4$ & $\star$ \\
1.43 & 177 & $\pm 14.2$ & $\star$ \\
1.53 & 140 & $\pm 14$ & $\star$ \\
1.75 & 202 & $\pm 40.4$ & $\star$ \\
\hline\hline
\end{tabular}
\caption{Observational $H(z)$ data (OHD). The data marked with stars is from Simon {\it et al.} sample, and the data marked with diamonds is
from Ruth {\it et al.} sample.} \label{tab:hz}
\end{center}
\end{table}
\begin{figure}
\centerline{\psfig{figure=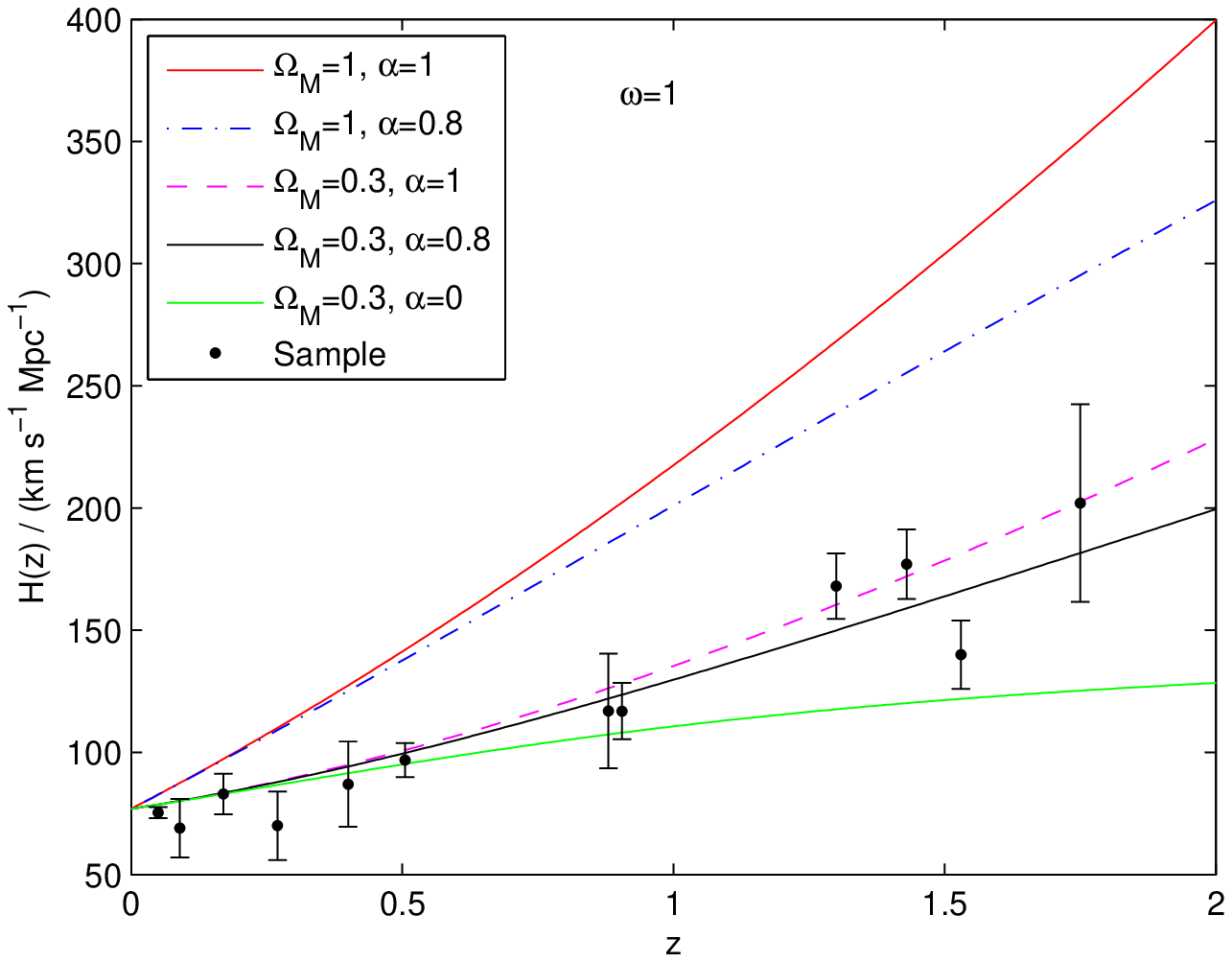,width=3.8truein,height=3.2truein} \hskip 0.0in} \caption{(color online). Hubble parameter $H(z)$ as a function
of redshift $z$ for a flat $\Lambda$CDM model with selected values of $\Omega_{\rm M}$ and $\alpha$. The data set in Table. \ref{tab:hz} are
also shown.} \label{fig:hz}
\end{figure}
\begin{figure}
\centerline{\psfig{figure=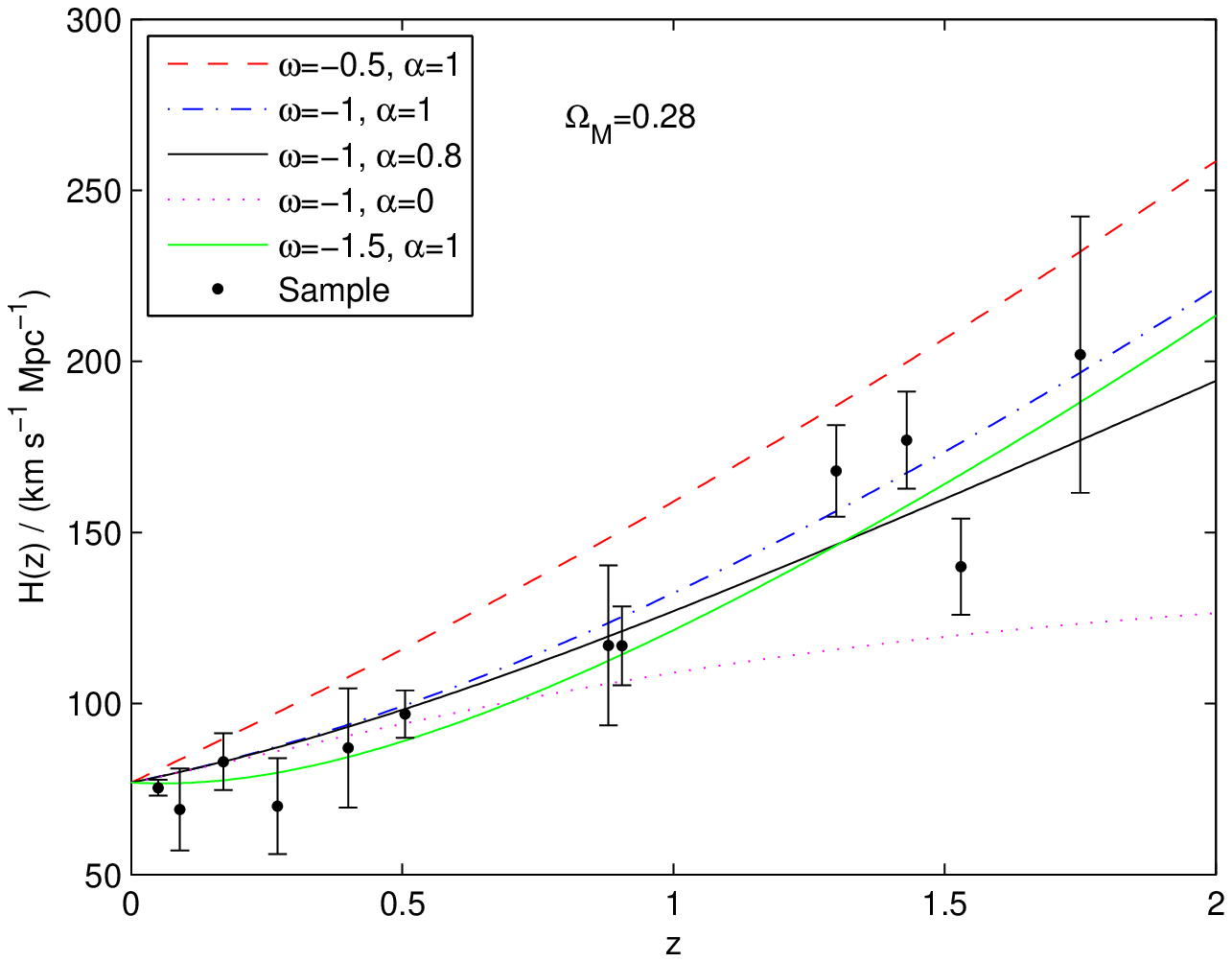,width=3.8truein,height=3.2truein} \hskip 0.0in} \caption{(color online). Hubble parameter $H(z)$ as a function
of redshift $z$ for a flat XCDM model with selected values of $\omega$ and $\alpha$. The data set in Table. \ref{tab:hz} are also shown.}
\label{fig:hzw}
\end{figure}
Considering Eq. (\ref{eq:DR}) again, we may change the variable by substituting redshift $z$ for the affine parameter $v$ and obtain,
\begin{equation}
(\frac{{\rm d}z}{{\rm d}v})^2\frac{{\rm d}^2D_{\rm A}}{{\rm d}z^2}+\frac{{\rm d}^2z}{{\rm d}v^2}\frac{{\rm d}D_{\rm A}}{{\rm
d}z}+\frac{2}{3}\alpha\Omega_{\rm M}(1+z)^5D_{\rm A}=0.
\end{equation}

Note that the universe discussed is spacially flat, i.e. $\Omega_{\rm k}=0$, so that $\Omega_{\rm \Lambda}=1-\Omega_{\rm M}$. Finally, after the
substitution of variable (\cite{Demianski}), we get a second-order ordinary differential equation in which the angular diameter distance $D_{\rm
A}$ is the function of redshift $z$, and $D_{\rm A}$ is in the unit of $c/H_0$:
\begin{equation}
    \frac{{\rm d}^2D_{\rm A}}{{\rm d}z^2}+{\cal P}\frac{{\rm d}D_{\rm A}}{{\rm d}z}+{\cal
    Q}D_{\rm A}=0,
    \label{eq:extendedDR}
\end{equation}
Here the initial conditions
\begin{equation}
    \left \{\begin{array}{ll}
    D_{\rm A}(0)=0,\\
    \displaystyle{\left. \frac{{\rm d}D_{\rm A}}{{\rm d}z} \right|_{z=0}=1.}
    \end{array}
    \right.
    \label{eq:extendedDRinitial}
\end{equation}
is satisfied.

Besides, the functions ${\cal P}$ and ${\cal Q}$ read
\begin{equation}
    \begin{array}{ll}
  \displaystyle{{\cal P}=\frac{\frac{7}{2}\Omega_{\rm M}(1+z)^3 + \frac{3\omega+7}{2}(1-\Omega_{\rm M})(1+z)^{3\omega+3}}
  {\Omega_{\rm M}(1+z)^4 + (1-\Omega_{\rm M})(1+z)^{3\omega+4}},}\\
    \displaystyle{{\cal Q}=\frac{\frac{3}{2}\alpha\Omega_{\rm M}+\frac{3\omega+3}{2}(1-\Omega_{\rm M})(1+z)^{3\omega}}
  {\Omega_{\rm M}(1+z)^2 + (1-\Omega_{\rm M})(1+z)^{3\omega+2}}}.
 \end{array}
\end{equation}
The numerical results of $D_{\rm A}$ and $\displaystyle{\frac{{\rm d}D_{\rm A}}{{\rm d}z}}$ (hereafter $D'_A(z)$) are shown in Fig.\ref{fig:ang}
with iterative calculations by the fourth-order Runge-Kutta scheme(see Sec.\ref{subsec:1}). From the well known Etherington principle---the
relation between the luminosity distance and angular diameter distance (\cite{Ether33}),
\begin{equation}
    D_{\rm L}=(1+z)^2D_{\rm A},
\end{equation}
we can get the luminosity distance as a function of $z$.

\begin{figure*}
\centerline{\psfig{figure=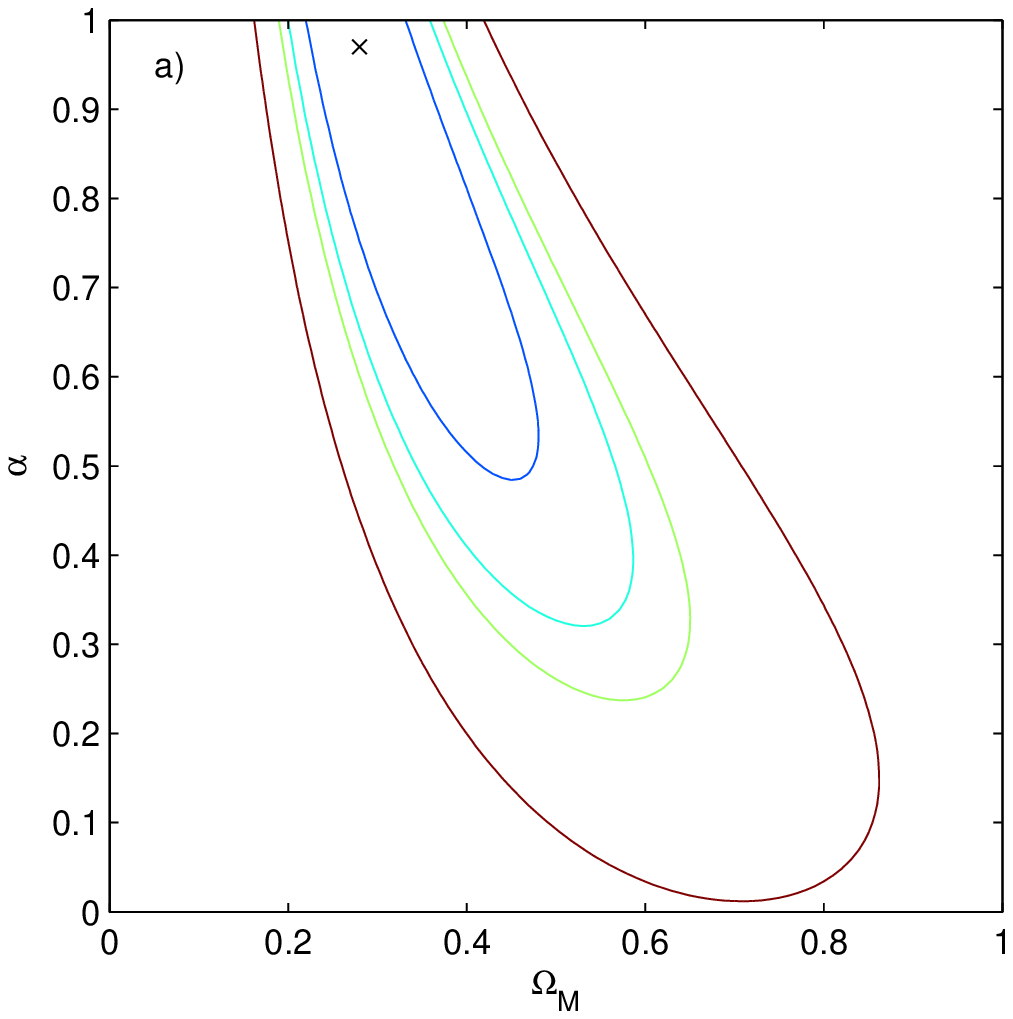,width=2.5truein,height=2.4truein} \psfig{figure=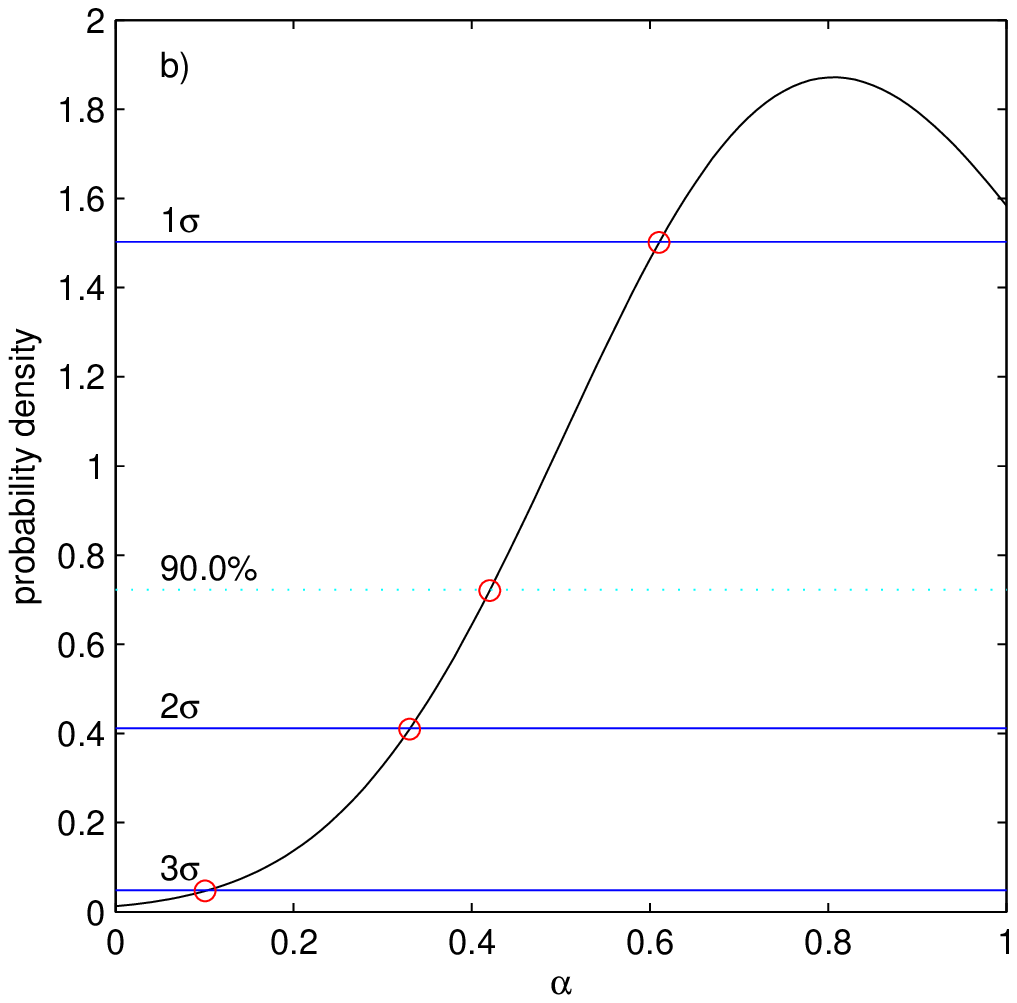,width=2.5truein,height=2.4truein}
\psfig{figure=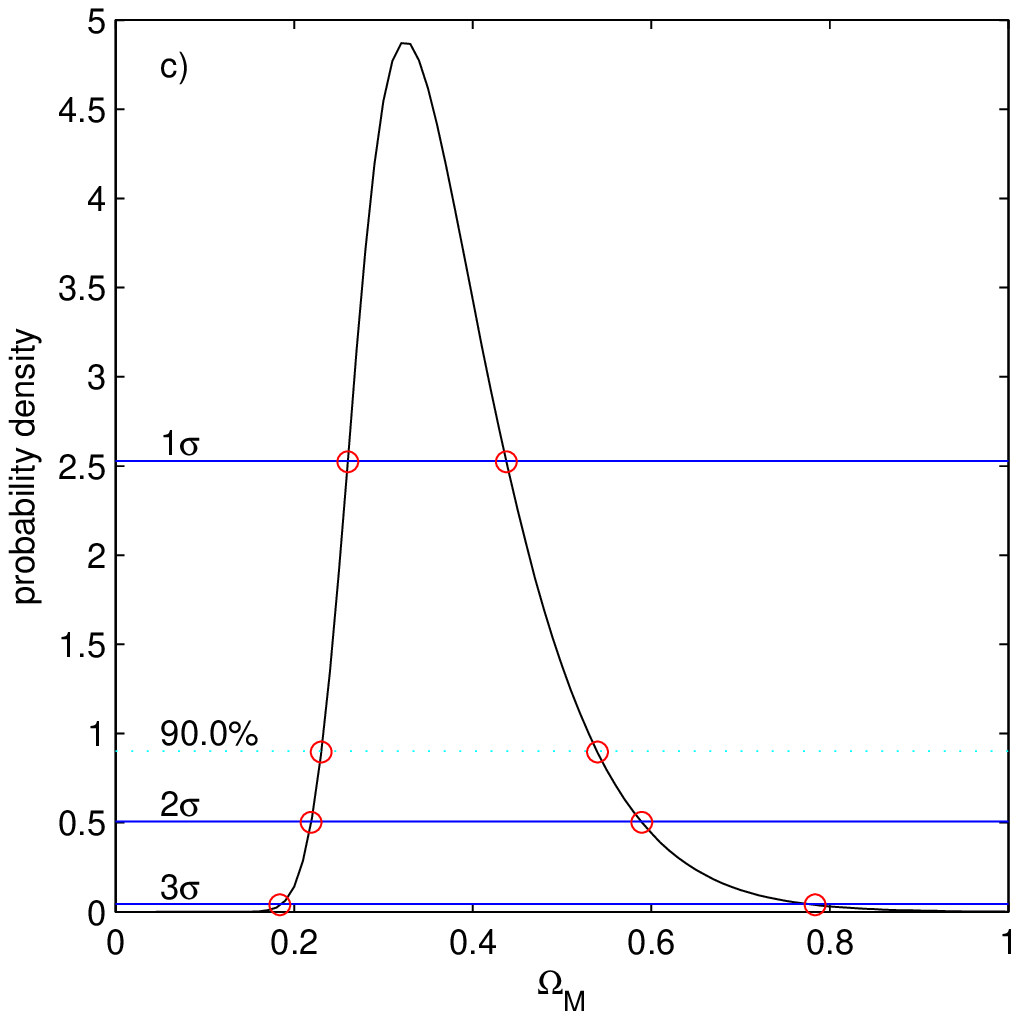,width=2.5truein,height=2.4truein}\hskip 0.0in} \caption{(color online). \textbf{(a)} Confidence regions at 68.3\%, 90.0\%,
95.4\%, and 99.7\% levels from inner to outer respectively on the ($\Omega_{\rm M}, \alpha$) plane for a flat $\Lambda$CDM model (without
considering the prior on $\Omega_{\rm M}$). The ``$\times$" in the center of confidence regions indicates the best-fit values (0.28,0.97).
\textbf{(b)} The one-dimensional probability distribution function (PDF) $p$ for the $\alpha$ parameter. \textbf{(c)} PDF for the $\Omega_{\rm
M}$ parameter.} \label{fig:contour}
\end{figure*}

\begin{figure*}
\centerline{\psfig{figure=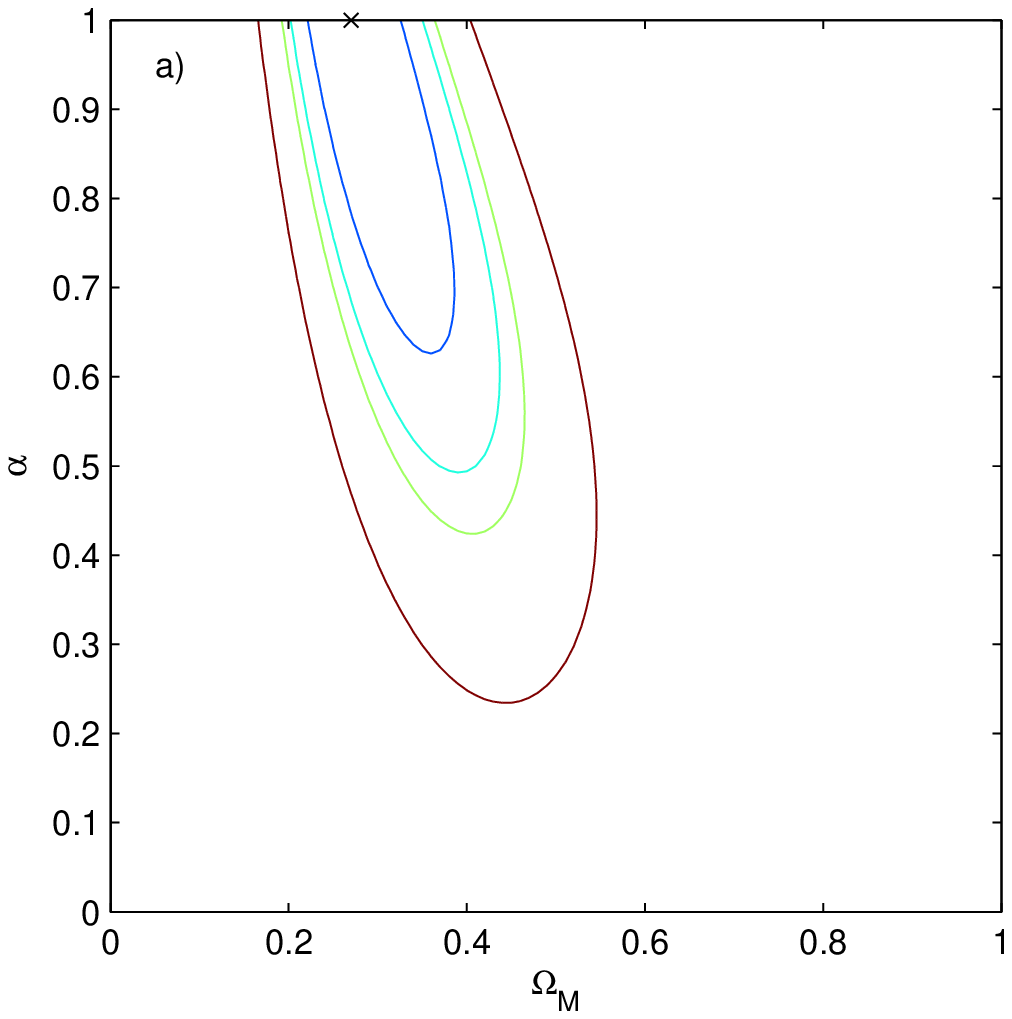,width=2.5truein,height=2.5truein} \psfig{figure=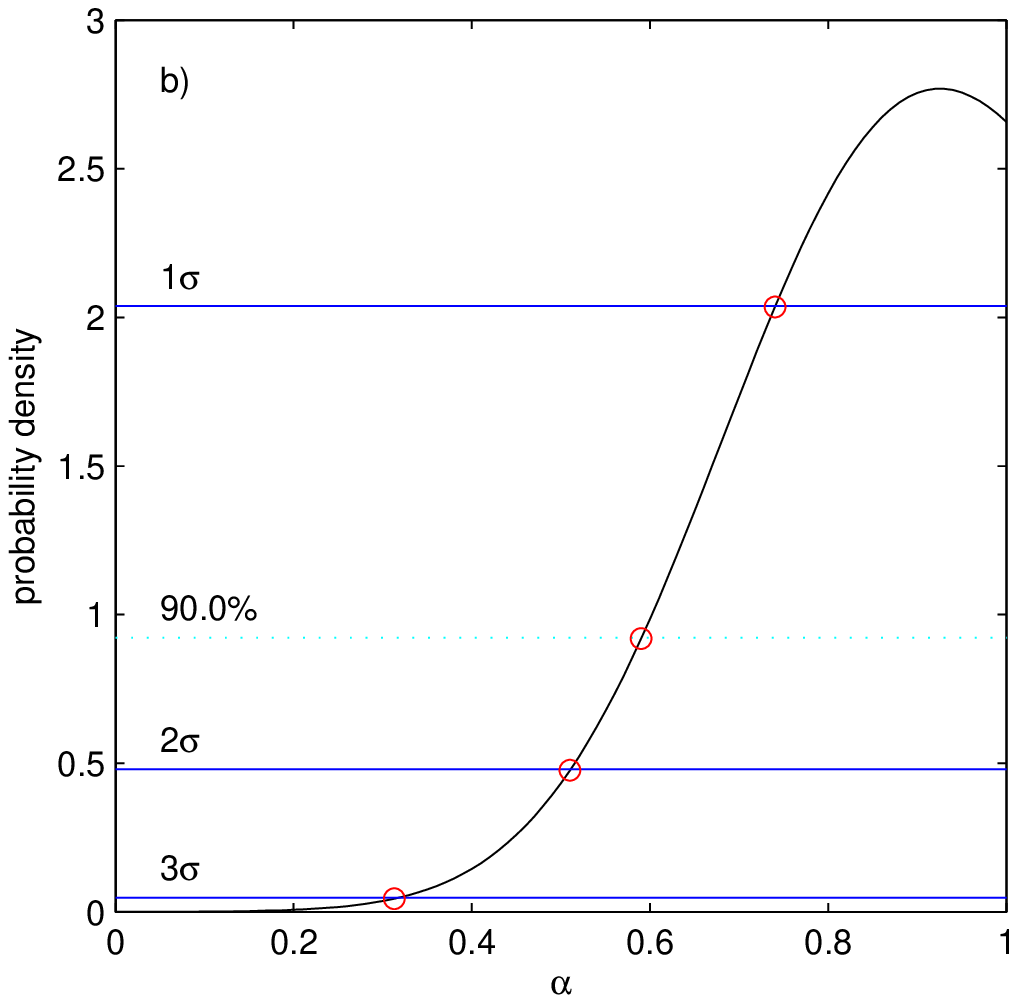,width=2.5truein,height=2.5truein}
\psfig{figure=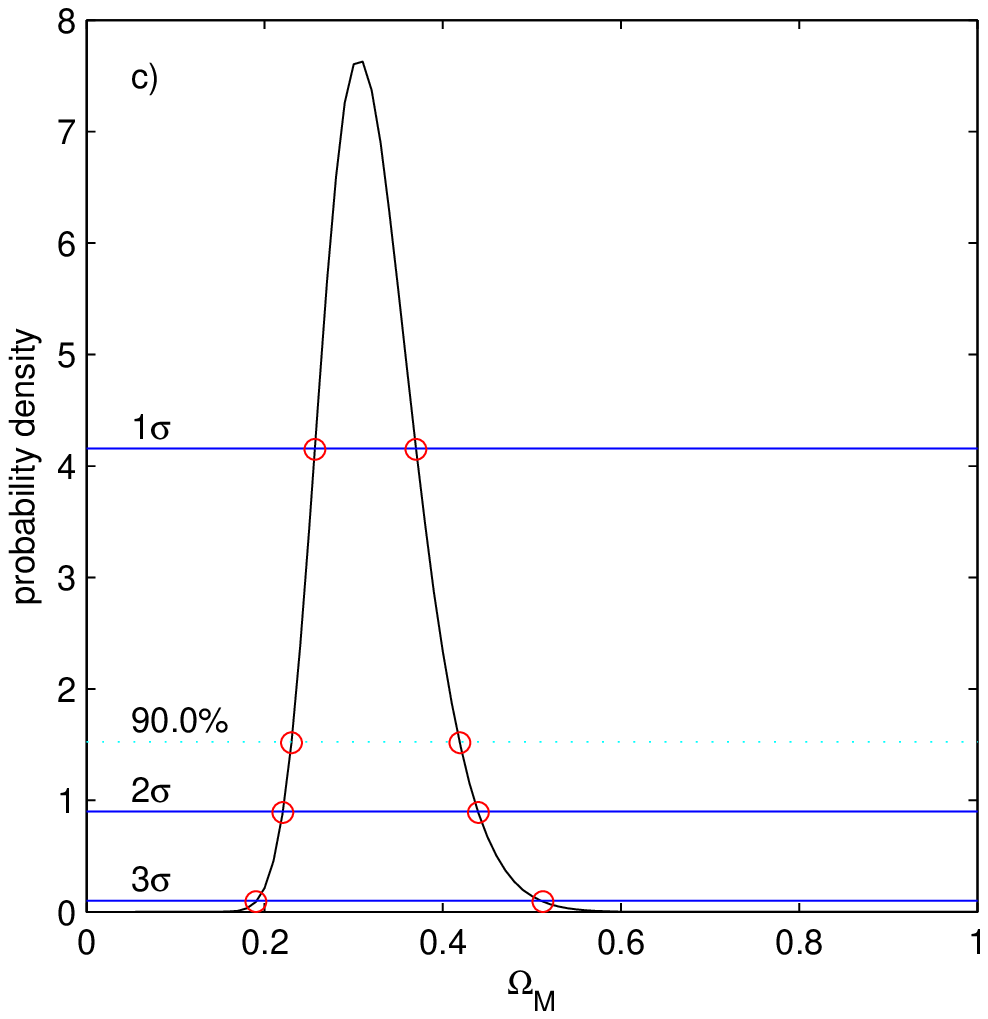,width=2.5truein,height=2.5truein} \hskip 0.0in} \caption{(color online). \textbf{(a)} Confidence regions at 68.3\%, 90.0\%,
95.4\%, and 99.7\% levels from inner to outer respectively on the ($\Omega_{\rm M}, \alpha$) plane for a flat $\Lambda$CDM model (a prior on
$\Omega_{\rm M}$ considered). The ``$\times$" in the center of confidence regions indicates the best fit values (0.27,1.0). \textbf{(b)} PDF for
the $\alpha$ parameter. \textbf{(c)} PDF for the $\Omega_{\rm M}$ parameter.}\label{fig:contourp}
\end{figure*}

\begin{figure*}
\centerline{\psfig{figure=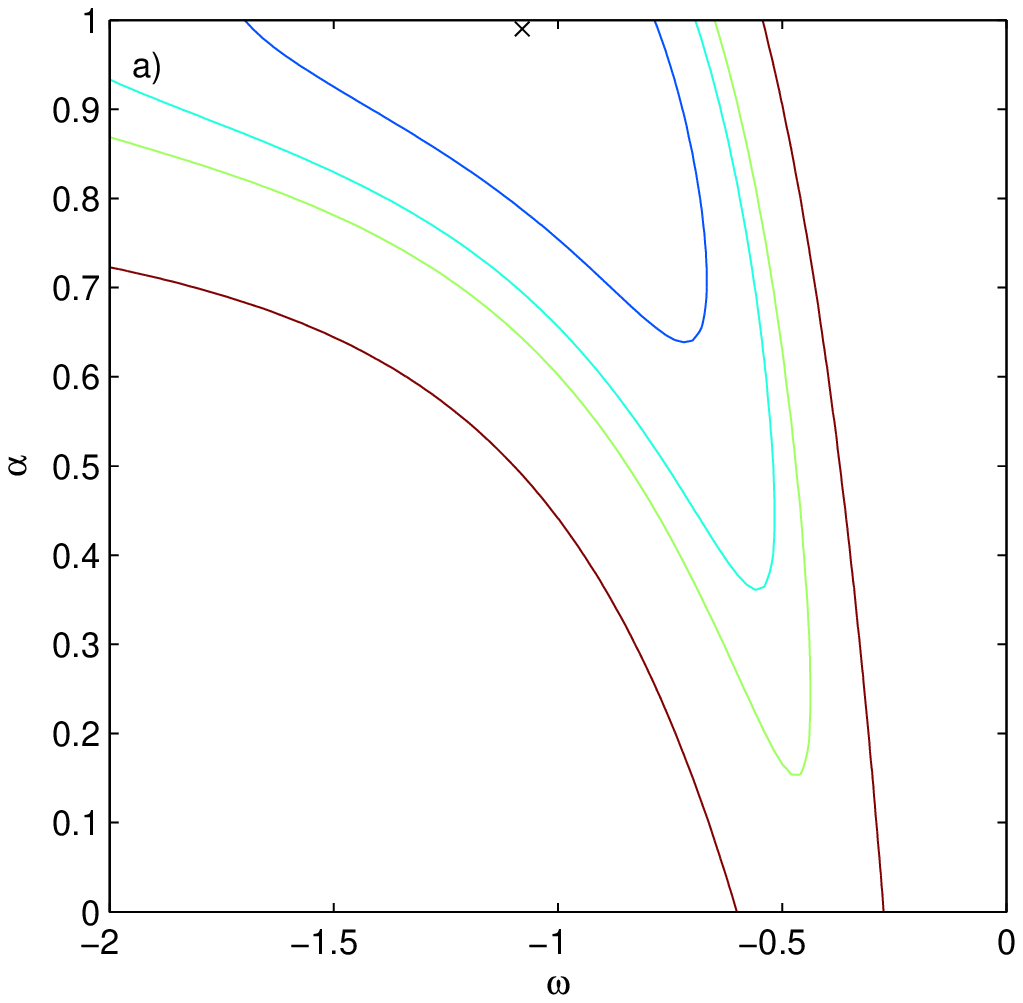,width=2.5truein,height=2.4truein} \psfig{figure=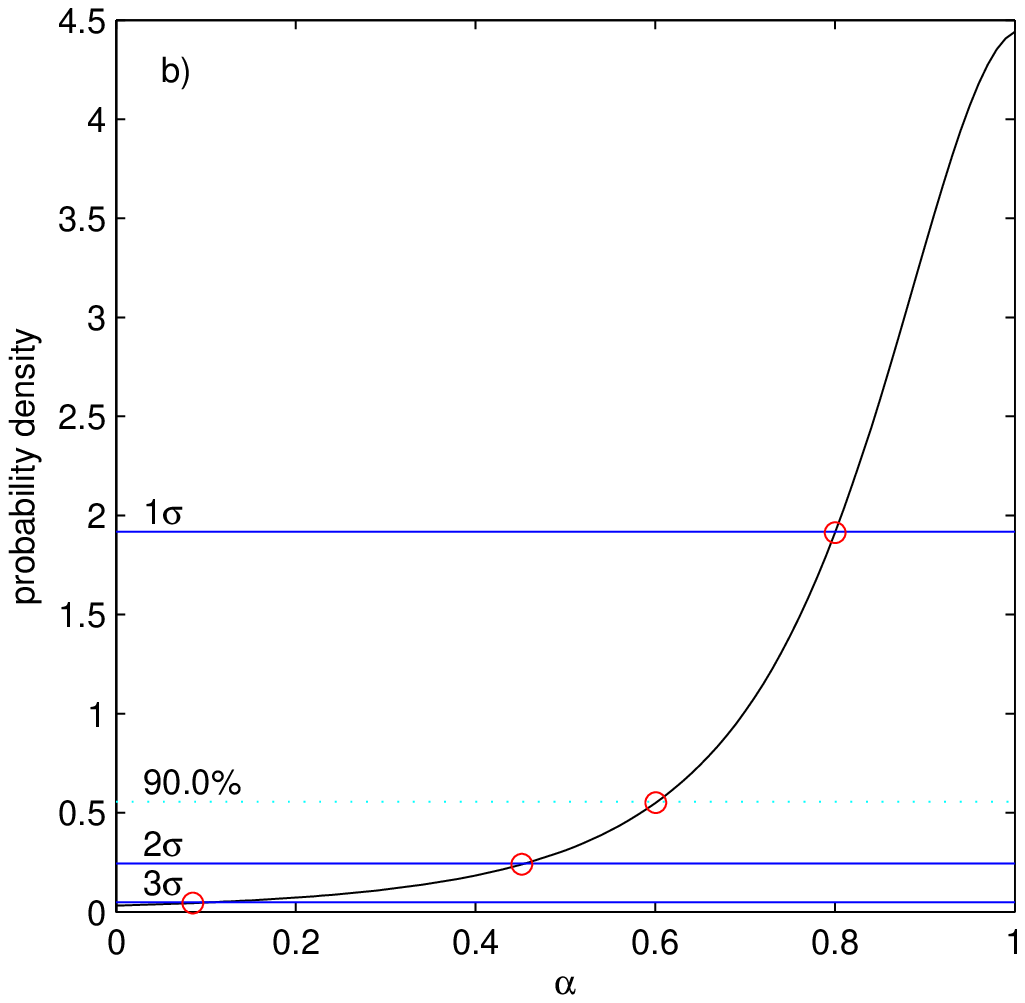,width=2.5truein,height=2.4truein}
\psfig{figure=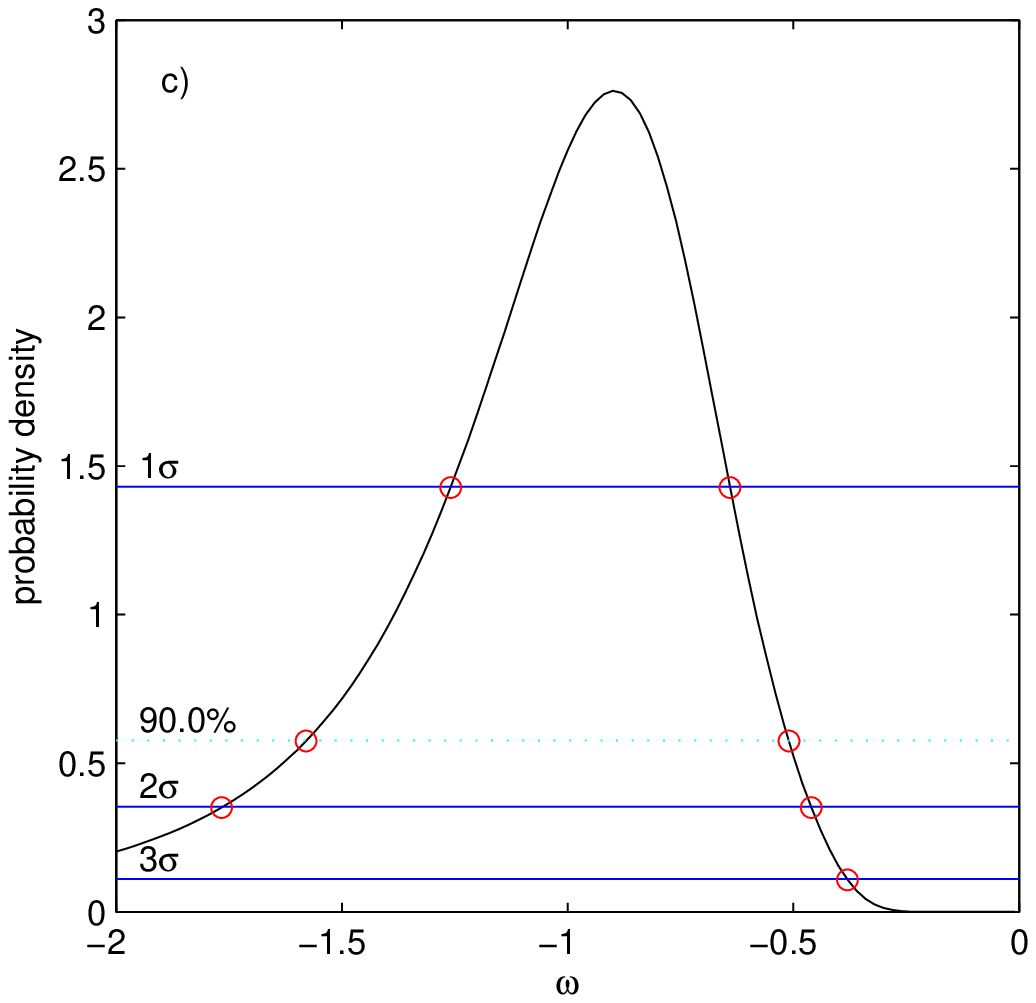,width=2.5truein,height=2.4truein}\hskip 0.1in} \caption{(color online). \textbf{(a)} Confidence regions at 68.3\%, 90.0\%,
95.4\%, and 99.7\% levels from inner to outer respectively on the ($\omega, \alpha$) plane for a flat XCDM model. The ``$\times$" in the center
of confidence regions indicates the best fit values (-1.08,0.99). \textbf{(b)} PDF for the $\alpha$ parameter. \textbf{(c)} PDF for the $\omega$
parameter.} \label{fig:contourw}
\end{figure*}
At present, by the aid of the method based on the differential age of the oldest galaxies, the Hubble parameter can be determined as a function
of redshift. It reads:
\begin{equation}
    H(z)=-\frac{{\rm d}z}{{\rm d}t}\frac{1}{1+z},
\end{equation}
which can be measured directly by the determination of ${\rm d}z/{\rm d}t$.

Since the comoving radial distance $r(z)$ (in units of $c/H_0$) in flat geometry can be expressed as
\begin{equation}
    r(z)=\int^z_0\frac{{\rm d}z}{E(z)},\label{eq:int}
\end{equation}
where $E(z)$ is the expansion rate of the universe, which relates Hubble parameter to Hubble constant $H_0$ in the equation:
\begin{equation}
    H(z)=H_0E(z),
    \label{eq:Hz}
\end{equation}
the angular diameter distance can be written as,
\begin{equation}
    D_{\rm A}=\frac{r(z)}{1+z}.\label{eq:dar}
\end{equation}
Differentiating Eq. (\ref{eq:int}) with respect to redshift $z$ and combining with Eq. (\ref{eq:dar}), we can get the expansion rate of the
universe expressed by $D_{\rm A}$ and $D'_A(z)$ at any redshift $z$:
\begin{equation}
    E(z)=\frac{1}{(1+z)D'_A(z)+D_{\rm A}}.
    \label{eq:Ez-Ra}
\end{equation}

\section{Samples and results}\label{sec:results}

\subsection{The observational data of $H(z)$}\label{subsec:1}

In order to constrain smoothness parameter and other cosmological parameters with OHD, we need to integrate Eq. (\ref{eq:extendedDR}) to obtain
$D_{\rm A}(z)$ and $D'_A(z)$ as a function of $z$ (Fig.\ref{fig:ang}), then from Eq. (\ref{eq:Hz}) and Eq. (\ref{eq:Ez-Ra}) to get Hubble
parameter as a function of redshift $z$ theoretically.

Although it is unable to obtain the analytical solution of Eq. (\ref{eq:extendedDR}) (\cite{Kantowski};\cite{Kantowski00};\cite{Kantowski01}),
one can get an approximate expression of the equation which is accurate enough to use in practical (\cite{Demianski}). It is convenient for
controlling the precision that we integrate the Eq. (\ref{eq:extendedDR}) iteratively applying the 4th-order Runge-Kutta scheme and then get
numerical results of $D_{\rm A}(z)$ and $D'_A(z)$. Furthermore, the Hubble parameter $H(\Omega_{\rm M}, \omega, \alpha, h; z)$ can be obtained
with any cosmological model at arbitrary $z$.

Note that OHD, consisting of two parts, is obtained from two different sources. For one part, from Simon {\it et al.} sample (\cite{Simon}), we
have a sample of 9 bins and $z\in[0, 1.75]$, while for the other, from Ruth {\it et al.} sample (\cite{Ruth}), we choose the data separated into
3 bins. In Table.\ref{tab:hz} we listed all the data and errors mentioned above with their sources marked, and we plotted them in
Fig.\ref{fig:hz} and Fig.\ref{fig:hzw} respectively. We can compare the theretical curves of $H(z)$ and the data in different models (see
Sec.\ref{subsec:2} and Sec.\ref{subsec:3}).

\subsection{Constraining $\alpha$ and $\Omega_{\rm M}$ for $\Lambda$CDM model}\label{subsec:2}

Firstly, we can find the relationship between $H(z)$ and $\alpha$, $\Omega_{\rm M}$. In Fig.\ref{fig:hz}, we plot the theoretical $H(z)$ at
redshift $z\in[0, 2]$ according to the last section for some typically selected $\alpha$ and $\Omega_{\rm M}$, where a flat $\Lambda$CDM model
is assumed. The figure include the mean values of OHD in redshift bins and their error bars. One can see from Fig.\ref{fig:hz} that the curve of
$H(z)$ strongly depends on $\Omega_{\rm M}$, i.e., the larger for $\Omega_{\rm M}$, the faster for the growth of $H(z)$. In contrast, the
smoothness-parameter $\alpha$ has effects on the properties of the Hubble parameter mainly at high redshift. For a certain $\Omega_{\rm M}$,
theoretical $H(z)$ with different $\alpha$ seem to appear similar at lower redshift, but they start to differ at high redshift.

In order to constrain $\alpha$ and $\Omega_{\rm M}$ we use $\chi^2$ minimization
\begin{equation}
    \chi^2(H_0,\alpha,\Omega_{\rm
    M})=\sum_{i=1}^{12}\left[\frac{H(H_0,\alpha,\Omega_{\rm M};z_i)-H_{\rm
    obs}(z_i)}{\sigma(z_i)}\right]^2,
\end{equation}
where $H(H_0,\alpha,\Omega_{\rm M};z_i)$ is the theoretical expecting of Hubble parameter which is determined by Eq. (\ref{eq:extendedDR}), Eq.
(\ref{eq:Ez-Ra}) and Eq. (\ref{eq:Hz}), and $H_{\rm obs}(z_i)$ is the observational values of the Hubble parameter with errors $\sigma(z_i)$ in
the sample.

In the analysis, we marginalize Hubble constant $H_0$ by integration over it, and assumed a Gaussian prior according to the best fitting value
obtained from Ref.(\cite{Bonamente}), i.e., $H_0=76.9^{+3.9}_{-3.4}$km ${\rm s}^{-1}$ ${\rm Mpc}^{-1}$. On the basis of the cosmic concordance
from observations, we can choose to add a Gaussian prior on $\Omega_{\rm M}$ optionally, i.e., $\Omega_{\rm M}=0.26\pm0.1$. We investigate the
minimization both considering this prior(Fig.\ref{fig:contourp}(a)) and without considering it(Fig.\ref{fig:contour}(a)). In
Fig.\ref{fig:contour}(a) and Fig.\ref{fig:contourp}(a) we plot the regions of confidence on the $\Omega_{\rm M}$ - $\alpha$ plane. The contours
on the confidence of 68.3\%, 90.0\%, 95.4\% and 99.7\% are determined by two-parameter levels 2.30, 4.61, 6.17 and 11.8, respectively.

We also plot the one-dimensional probability distrinbution functions $p$ (PDF) of parameters $\Omega_{\rm M}$ and $\alpha$. In
Fig.\ref{fig:contour}(b) and Fig.\ref{fig:contour}(c), PDFs were plotted without considering the Gaussian prior, while in
Fig.\ref{fig:contourp}(b) and Fig.\ref{fig:contourp}(c) we considered this prior. One can see that in 90\% confidence region,
$0.59\leq\alpha\leq1.0$ and $0.23\leq\Omega_{\rm M}\leq0.41$ if we consider the prior. While without any prior a 90\% confidence lies in the
region of $0.42\leq\alpha\leq1.0$ and $0.23\leq\Omega_{\rm M}\leq0.54$. The ``$\times$"s indicate the model with the best fitting values that
occur at $\alpha=1$, $\Omega_{\rm M}=0.27$ and $\alpha=0.97$ , $\Omega_{\rm M}=0.28$, respectively. It is clear that whether or not we consider
the Gaussian prior, the best fitting models are nearly the same with $\alpha$ slightly lower than one, corresponding to a universe with the
uniform distribution of cold dark matter, and the value of $\Omega_{\rm M}$ favors other observations.

\subsection{Constraining $\alpha$ and $\omega$ for XCDM model}\label{subsec:3}

In this part, we constrain $\alpha$ and $\omega$ for XCDM model. In this case, we set the density-parameter $\Omega_{\rm M}$ to be 0.28 from the
best fitting results in the Sec.\ref{subsec:2}. In Fig.\ref{fig:hzw} we plot theoretical $H(z)$ with some typically selected $\alpha$ and
$\omega$, from which one can see how these two parameters modify the theoretical curve. Unfortunately, we find that the curves is not so
strongly denpend on the parameters as in Sec.\ref{subsec:2}. We constrain these two parameters by $\chi^2$ minimization
\begin{equation}
    \chi^2(H_0,\alpha,\omega)=\sum_{i=1}^{12}\left[\frac{H(H_0,\alpha,\omega;z_i)-H_{\rm
    obs}(z_i)}{\sigma(z_i)}\right]^2,
\end{equation}
where $H(H_0,\alpha,\omega;z_i)$ is the theoretical value of Hubble parameter while $H_{\rm obs}(z_i)$ and $\sigma(z_i)$ are their observational
values and errors, respectively. Again, marginalizing the parameter $H_0$, we get the regions of confidence on the $\omega$ - $\alpha$ plane.
Then integrating over $\omega$ and $\alpha$ in the two-dimensional probability function $p(\alpha,\omega)$, we obtain PDFs of $\alpha$ and
$\omega$. The numerical results were plotted in Fig.\ref{fig:contourw}.

From Fig.\ref{fig:contourw}(a), we find $\alpha$ is only mildly constrained. The best fitting point occurs at $\alpha=0.99$ and $\omega=-1.08$
which indicates that the state of equation of XCDM is approximately that of the cosmological constant. All the allowable values of $\alpha$ in
$[0,1]$ are permitted at the $3\sigma$ confidence level.

\section{Conclusions and Discussions}
In this article, we study how the inhomogeneous distribution of cold dark matter affects the Hubble parameter at different redshift. By using
OHD from Simon {\it et al.} together with Ruth {\it et al.} in a flat $\Lambda$CDM model, the smoothness-parameter and density-parameter are
constrained at different confidence intervals. By marginalizing the Hubble constant $H_0$, we found that the best fitting values are
$\alpha=0.97$ and $\Omega_{\rm M}=0.28$ for $\Lambda$CDM model. While if we assume a Gaussian prior of $\Omega_{\rm M}=0.26\pm0.1$, we found
that the best fitting values are $\alpha=1$ and $\Omega_{\rm M}=0.27$. With the XCDM model, setting $\Omega_{\rm M}=0.28$, we get the best
fitting values $\alpha=0.99$ and $\omega=-1.08$. Comparing with the constraints on the smoothness-parameter with samples of compact radio
sources --- all the values of $\alpha$ from 0 to 1 are allowed at 68.3\% statistical confidence level (\cite{SantosR}), the constraint on
luminosity distance from SNe Ia gave little better results, which is $0.42\leq\alpha\leq1.0$ and $0.25\leq\Omega_{\rm M}\leq0.44$ at 90\%
confidence level (\cite{SantosSN}). In our work, from Fig.\ref{fig:contour} and Fig.\ref{fig:contourp}, the empty beam ($\alpha=0$) case is
excluded at even $3\sigma$ confidence level. This result is better than the two previous works (\cite{SantosSN,SantosR}) which constrain
$\alpha$ parameter. We can see that OHD constrains the smoothness parameter more effectively than both SNe Ia data and the data of angular
diameter of compact radio sources.

In the meanwhile, as one can see from Fig.\ref{fig:hz} and Fig.\ref{fig:hzw}, the errors of the data is too large to constrain $\alpha$ into a
relatively small interval, especially at high redshift, or to let $\alpha$ be a somehow accurate value. The statistical effects may not be
neglected in this analysis. In the near future, we expect better constraint on $\alpha$ from more accurate data and/or more data at high
redshift. In addition, OHD, also angular diameter distances of compact radio sources and luminosity distances of SNe Ia, may also enable us to
study the $\alpha(z)$ as a function of $z$.

Throughout our work, we discuss the smoothness-parameter independent of space-time, i.e., $\rho_{\rm int}+\rho_{\rm clus}$ is constant
everywhere, and $\alpha$ only describes the ratio of them. In a real universe the structures of walls and voids could not be described by such
model, therefore these complex structures, which might be characterized by more parameters and/or more complicated models, may need more and
more precise data. Paul Hunt {\it et al.} have discussed the case that we located in a 200-300 Mpc void with terribly low density, which is
expanding at a rate 20-30\% higher than the average rate (\cite{Hunt}). In the future work, we should consider more complex models better
describe our universe, together with consummate physical theories, in order to find out more accurate answer. \normalem

\begin{acknowledgements}
H.R.Y. would like to thank Jing Wang, Zhong-Xu Zhai and Cheng-Xiao Jiang for their kindly help. This work was supported by the National Science
Foundation of China (Grants No.10473002), the Ministry of Science and Technology National Basic Science program (project 973) under grant
No.2009CB24901, Scientific Research Foundation of Beijing Normal University and the Scientific Research Foundation for the Returned Overseas
Chinese Scholars, State Education Ministry.
\end{acknowledgements}


\begin{thebibliography}{99}
\small \setlength{\itemindent}{-3mm} \setlength{\itemsep}{-0.5mm} \setlength{\baselineskip}{4.5mm}

\bibitem[{Perlmutter} {et~al.}(1998)]{perm1998} S. Perlmutter et al., 1998, Nature 391, 51

\bibitem[{Perlmutter} {et~al.}(1999)]{perm1999} S. Perlmutter et al., 1999, ApJ, 517, 565

\bibitem[{Riess} {et~al.}(1998)]{Riess} A. G. Riess et al., 1998, AJ, 116, 1009

\bibitem[{Riess} {et~al.}(2007)]{Riess07} A. G. Riess  et al., 2007, ApJ 659, 98

\bibitem[{Efstathiou} {et~al.}(2002)]{Ef02} G. Efstathiou et. al., 2002, MNRAS, 330, L29

\bibitem[{Allen} {et~al.}(2004)]{Allen04} S. W. Allen et al., 2004, MNRAS, 353, 457

\bibitem[{Astier} {et~al.}(2006)]{Astier06} P. Astier et al., 2006, A \& A, 447, 31

\bibitem[{Spergel} {et~al.}(2002)]{Spergel02} D. N. Spergel et al., 2007, ApJS, 170, 377

\bibitem[{Carrllo} {et~al.}(1992)]{Carrllo} S. M. Carrllo et al., 1992, preprint (hep-th/9207037)

\bibitem[{Caldwell} {et~al.}(1998)]{Caldwell98} R. R. Caldwell et al., 1998, PRL, 80, 1582

\bibitem[{Kamenshchik} {et~al.}(2000)]{Kam00} A. Y. Kamenshchik et al., 2000, PLB, 487, 7

\bibitem[{Turner} {et~al.}(1997)]{turner97} M. S. Turner \& M. White, 1997, PRD, 56, R4439

\bibitem[{Chiba} {et~al.}(1997)]{Chiba97} T. Chiba, N. Sugiyama \& T. Nakamura, 1997, MNRAS, 289, L5

\bibitem[{Alcaniz} {et~al.}(1997)]{Alcaniz99} J. S. Alcaniz \& J. A. S. Lima, 1999, ApJ, 521, L87

\bibitem[{Alcaniz} {et~al.}(2001)]{Alcaniz01} J. S. Alcaniz \& J. A. S. Lima, 2001, ApJ, 550, L133

\bibitem[{Lima} {et~al.}(2000)]{Lima00} J. A. S. Lima \& J. S. Alcaniz, 2000, MNRAS, 317, 893

\bibitem[{Lima} {et~al.}(2003)]{Lima03} J. A. S. Lima, J. V. Cunha \& J. S. Alcaniz, 2003, PRD, 68, 023510

\bibitem[{D\'{a}browski} {et~al.}(2007)]{browski07} M. P. D\'{a}browski, 2007, preprint (arXiv:gr-qc/0701057)

\bibitem[{Kamenshchik} {et~al.}(2000)]{Csaki} C. Csaki et al., 2000, PRD, 62, 045015

\bibitem[{Freese} {et~al.}(2002)]{Freese} K. Freese \& M. Lewis, 2002, PLB, 540, 1

\bibitem[{Weinberg} {et~al.}(1989)]{Weinberg89} S. Weinberg, 1989, RMPh, 61, 1

\bibitem[{Zeldovich} {et~al.}(1967)]{Zel67} Zeldovich, Ya. B. 1967, Pis'ma Zh. Eksp. Teor. Fiz., 6, 883 (1967, JETP. Lett., 6, 316)

\bibitem[{Dashveski} {et~al.}(1966)]{Dashv66} Dashveski, V. M., Slysh, V. I., 1966, Soviet Astr., 8, 854

\bibitem[{Kayser} {et~al.}(1997)]{Kayser97} Kayser, R. et al. 1997, A \& A, 318, 680

\bibitem[{Dyer} {et~al.}(1973)]{DR73} C. C. Dyer \& R. C. Roeder, 1973, ApJ, 180, L31-L34

\bibitem[{Dyer} {et~al.}(1972)]{DR72} C. C. Dyer \& R. C. Roeder, 1972, ApJ, 174, L115-L117

\bibitem[{Santos} {et~al.}(2008)]{SantosSN} R. C. Santos, J. V. Cunha, \& J. A. S. Lima, 2008, PRD, 77, 023519

\bibitem[{Santos} {et~al.}(2008)]{Astier} P. Astier et al., 2006, A \& A, 447, 31

\bibitem[{Riess} {et~al.}(2007)]{Riess2} A. G. Riess et al., 2007, ApJ, 659, 98

\bibitem[{Santos} {et~al.}(2008)]{SantosR} R. C. Santos \& J. A. S. Lima, 2008, PRD, 77, 083505

\bibitem[{Yi} {et~al.}(2007)]{Yi} Ze-Long Yi \& Tong-Jie Zhang, 2007, MPhLA, 22, 01

\bibitem[{Wan} {et~al.}(2007)]{Wan} Hao-Yi Wan et al., 2007, PLB, 651, 5-6

\bibitem[{Lin} {et~al.}(2008)]{Lin} Hui Lin et al., 2008, arXiv:0804.3135v2

\bibitem[{Zhang} {et~al.}(2008)]{Zhanghongsheng}  Hong-Sheng Zhang \& Zong-Hong Zhu, 2008, JCAP, 03, 007

\bibitem[{Zhang} {et~al.}(1966)]{Sachs66} R. K. Sachs, J. Kristian, 1966, ApJ, 143, 379

\bibitem[{Demianski} {et~al.}(2003)]{Demianski} M. Demianski et al., 2003, A \& A, 411, 33-40

\bibitem[{Sachs} {et~al.}(1961)]{Sachs61} R. K. Sachs, 1961, PRSLA, 264, 309

\bibitem[{Schroedinger} {et~al.}(1956)]{Sch56} E. Schroedinger, 1956, Expanding Universes (Cambridge: Cambridge University Press) chap. 2

\bibitem[{Etherington} {et~al.}(1933)]{Ether33} I. M. H. Etherington, 1933, Phil. Mag., 15, 761

\bibitem[{Schneider} {et~al.}(1988a)]{Schneider} P. Schneider \& A. Weiss, 1988a, ApJ, 327, 526

\bibitem[{Schneider} {et~al.}(1988b)]{Schneider88} P. Schneider \& A. Weiss, 1988b, ApJ, 330, 1

\bibitem[{Bartelmann} {et~al.}(1991)]{Bartelmann91} M. Bartelmann \& P. Schneider, 1991, A \& A, 248, 349

\bibitem[{Watanabe} {et~al.}(1992)]{Watanabe92} K. Watanabe,  M. Sasaki \& K. Tomita, 1992, ApJ, 394, 38

\bibitem[{Kantowski} {et~al.}(1998)]{Kantowski} R. Kantowski, 1998, ApJ, 507, 483

\bibitem[{Kantowski} {et~al.}(2000)]{Kantowski00} R. Kantowski, J. K. Kao, R. C. Thomas, 2000, ApJ, 545, 549

\bibitem[{Kantowski} {et~al.}(2001)]{Kantowski01} R. Kantowski, R. C. Thomas, 2001, ApJ, 561, 491

\bibitem[{Simon} {et~al.}(2005)]{Simon} J. Simon et. al., 2005, PRD, 71, 123001

\bibitem[{Ruth} {et~al.}(2005)]{Ruth} A. D. Ruth et. al., 2008, ApJ, 677, 1-11

\bibitem[{Bonamente} {et~al.}(2005)]{Bonamente} M. Bonamente et. al., 2006, ApJ, 647, 25

\bibitem[{Hunt} {et~al.}(2008)]{Hunt} Paul Hunt et. al., 2008, preprint (arXiv:0807.4508)

\end{thebibliography}
\end{document}